\documentclass{ws-procs975x65}

\newcommand{\be}{\begin{equation}}
\newcommand{\ee}{\end{equation}}

\begin{document}

\title{\uppercase{Absence of scalar hair in scalar-tensor  
gravity}}

\author{\uppercase{Valerio Faraoni}}
\address{Physics Department and STAR Research Cluster, 
Bishop's University, 2600 College St., Sherbrooke, 
Qu\'{e}bec, Canada J1M~1Z7\\
E-mail: vfaraoni@ubishops.ca}
\author{\uppercase{Thomas P. Sotiriou}}
\address{SISSA, Via Bonomea 265, 34136 Trieste, Italy and INFN, Sezione di Trieste\\
E-mail: sotiriou@sissa.it}

\begin{abstract}
Stationary, asymptotically flat black holes in 
scalar-tensor theories of gravity are studied. It is 
shown that such black holes have no scalar hair and are the
same as in General Relativity. 
 \end{abstract}


\bodymatter

%%%%%%%%%%%%%%%%%%%%%%%%%%%%%55
\section{Introduction}
%\label{aba:sec1}

In General Relativity (GR) stationary black holes, which are the endpoint of 
gravitational  collapse, must be  axisymmetric and are 
described by the Kerr-Newman metric\cite{Hawking72b}. The 
prototypical alternative  theory of gravity is 
Brans-Dicke theory with (Jordan frame) action
\be
S_{BD}=\int d^4x \sqrt{-\hat{g}} \left[ \varphi 
\hat{R}-\frac{\omega_0}{\varphi} \hat{\nabla}^\mu \varphi 
\hat{\nabla}_\mu \varphi+L_m(\hat{g}_{\mu\nu},\psi)\right] \, .
\ee
In 1972 Hawking showed that stationary black holes  in this 
theory  must be the Kerr-Newman black holes of GR\cite{Hawking72}. This 
result was generalized by Bekenstein to more general 
scalar-tensor theories, but with the additional assumption of spherical symmetry\cite{Bekenstein96}. 
%When trying to extend Hawking's result, one is more or less restricted   
%to spherical symmetry and it is difficult to go beyond. In this context a new proof\cite{PRL}  
%seems significant because it is  particularly simple while being general. This work  
Hawking's result has recently been extended to  general scalar-tensor theories with action
\be
S_{ST}=\int d^4x \sqrt{-\hat{g}} \Big[\varphi 
\hat{R}-\frac{\omega(\varphi)}{\varphi} \, \hat{\nabla}^\mu \varphi 
\hat{\nabla}_\mu \varphi -V(\varphi) 
+L_m(\hat{g}_{\mu\nu},\psi)\Big] 
\ee
without any extra assumption of symmetry apart from 
stationarity\cite{PRL}. This proof is presented below.

\section{The proof}

To begin with, we require:

\begin{itemize}
\item {\bf Asymptotic flatness}: this requires $V(\varphi_0)=0$ and $
\varphi_0 \, V'(\varphi_0)=2V(\varphi_0) $, where $\varphi_0$ is the value the Brans-Dicke scalar field 
approaches  as 
$r\rightarrow +\infty$ (gravitational collapse occurs on scales 
much smaller than the Hubble scale $ H_{0}^{-1}$, so 
asymptotic flatness is expected to be an adequate 
approximation physically).  

\item {\bf Stationarity}: the black hole is supposed to be 
the endpoint of collapse.
\end{itemize}

We map the theory to the  Einstein conformal frame according to $
\hat{g}_{\mu\nu} \rightarrow  g_{\mu\nu}=\varphi \,\hat{g}_{\mu\nu} $, $
\varphi \rightarrow \phi $, with $
d\phi=\sqrt{\frac{2\omega(\varphi)+3}{16\pi}} \,\,  
\frac{d\varphi}{\varphi} $ (for $  \omega \neq -3/2  $).  The action becomes 
\be
S_{ST} = \int d^4x \sqrt{-g} \Big[\frac{R}{16\pi}-\frac{1}{2} 
\nabla^\mu \phi \nabla_\mu \phi 
-U(\phi)+L_m(\hat{g}_{\mu\nu},\psi)\Big] \,,
\ee
where $ U(\phi)=V(\varphi)/\varphi^2 $.  The  field 
equation for the scalar {\em in vacuo} in the Einstein frame is
\begin{equation}
\label{scalar}
\Box\phi=U'(\phi)\,.
\end{equation}
Since the conformal factor of the transformation  depends 
only on the Brans-Dicke field $\varphi$, the Einstein 
frame symmetries are the same as in the Jordan frame. In 
particular, there exists a Killing vector $\xi^{\mu}$ which 
is timelike at infinity (stationarity). In the Einstein 
frame and in electrovacuum the theory is essentially GR 
with a minimally coupled scalar field. So, stationary, 
asymptotically flat black holes have to be axisymmetric 
and, hence, there should be a second Killing  vector  
$\zeta^{\mu}$ which is spacelike at infinity, provided that 
the stress-energy tensor for $\phi$ satisfies the weak 
energy condition\cite{Hawking72b}. Consider, {\em in vacuo},  a 4-volume ${\cal V}$ bounded by the 
horizon $H$, two partial Cauchy hypersurfaces ${\cal S}_1$, 
${\cal S}_{2}$, and a  timelike 3-surface at infinity.  
Now multiply both sides of eq.~(\ref{scalar}) by $U'$ and integrate over the 
4-volume ${\cal V}$, obtaining
\be
\int_{\cal V} d^4 x\sqrt{-g}\,  U'(\phi)\Box \phi= 
\int_{\cal V}  d^4 x\sqrt{-g}\, U'^2(\phi) \,.
\ee
We can rewrite this equation as
\be
\int_{\cal V} d^4 x\sqrt{-g}\, \big[U''(\phi)\nabla^\mu 
\phi\nabla_\mu \phi + 
U'^2(\phi)\big]
=\int_{\partial{\cal V}} d^3 
x\sqrt{|h|}\, U'(\phi)  n^\mu \nabla_\mu\phi \,,
\ee
where $n^{\mu}$ is the normal to the boundary and $h$ is the determinant of the 
induced metric $h_{\mu\nu}$ on this boundary. Splitting the 
boundary into its constituent parts 
%yields
%\be
%\int_{\cal V}= \int_{{\cal S}_1} + \int_{{\cal S}_2} + 
%\int_{horizon} + \int_{r=\infty} 
%\ee
%Now 
one has $\int_{r=\infty}=0 $,
\begin{equation}
\int_{horizon} d^3  x\sqrt{|h|}\, U'(\phi)  n^\mu \nabla_\mu\phi 
=0\,,
\end{equation}
because of the spacetime symmetries, and $
\int_{{\cal S}_1}=-\int_{{\cal S}_2} $ if $S_2$ is obtained by shifting each point of $S_1$ along integral curves of $\xi^\mu$, hence
\be
 \int_{\cal V} d^4 x\sqrt{-g}\, \big[U''(\phi)\nabla^\mu 
\phi\nabla_\mu \phi + U'^2(\phi)\big]=0 \,. 
\ee
$U'^2 \geq 0$, $\nabla^{\mu} \phi$ (which is orthogonal to 
both $\xi^{\mu}$ and $ \zeta^{\mu}$) is spacelike or zero, 
and with 
$U''(\phi) \geq 0$ being a local stability condition, 
one concludes that it must be $\nabla_{\mu}\phi \equiv  0$ in ${\cal V}$ 
and $U'(\phi_0)=0$. But for  $\phi=$const., to which we have reduced, the scalar-tensor theory reduces 
to GR and the black hole must be described by the Kerr metric.

Metric $f(R)$ gravity, which has seen much recent attention\cite{Sotiriou:2008rp, review2}, is a special  
Brans-Dicke theory with parameter $\omega=0$ and a non-trivial potential $V$ for the Brans-Dicke field $\varphi=f'(R)$. 
Palatini $f(R)$ gravity, instead, corresponds to an $\omega=-3/2$ Brans-Dicke theory (again, with 
a potential). The case $\omega=-3/2$ was explicitly excluded in our discussion but
  $\omega=-3/2$ Brans-Dicke theory  reduces to GR {\em in vacuo} anyway.

\section{Conclusions}
The proof presented above extends immediately to electro-vacuum and to any form of conformal 
matter with trace of the energy-momentum tensor $T=0$.
It implies that asymptotically flat black holes that are the endpoint of collapse in scalar-tensor gravity are 
described by the Kerr-Newman metric.  The assumption of asymptotic flatness is 
a limitation mathematically, but one expects on physical grounds that the effect of a  
Friedmann-Lema\^itre-Robertson-Walker asymptotic structure on 
astrophysical  collapse to be completely negligible (except for  primordial black holes for which the collapse and the Hubble 
scales can be comparable\cite{Jacobson:1999vr}). 

There are certain exceptions to the proof, which include:
(i) theories in which $\omega \rightarrow \infty $ somewhere outside or on the horizon;
(ii)  theories in which $\varphi\rightarrow \infty $ or $\varphi\to 0$ somewhere outside or on the horizon; \footnote{An example of a solution where $\varphi\rightarrow \infty $ on the horizon is that of Bocharova {\em et al.}\cite{maverick}  (which is, however, unstable\cite{unstable}).}
(iii) theories in which the stress-energy tensor of the Einstein-frame scalar violates the weak energy condition.

It is likely that the majority of these exceptional 
theories or solutions will be unphysical ({\em e.g.}, the 
gravitational coupling in scalar-tensor gravity is 
inversely  proportional to $\varphi$) but interesting 
exceptions might exist.
This issue will be addressed in future work.

\section*{Acknowledgments}

We are grateful to V. Vitagliano and S. Liberati for 
discussions. VF acknowledges financial support by Bishop's 
University and {\em NSERC}. TPS acknowledges financial
support provided under the Marie Curie Career Integration
Grant 	LIMITSOFGR-2011-TPS and the European Union's FP7 ERC Starting Grant "Challenging General Relativity" CGR2011TPS, grant agreement no. 306425.

\bibliographystyle{ws-procs975x65}
\bibliography{ws-pro-sample}

\end{document}